\documentclass[sigconf,nonacm]{acmart}

\AtBeginDocument{%
  \providecommand\BibTeX{{%
    \normalfont B\kern-0.5em{\scshape i\kern-0.25em b}\kern-0.8em\TeX}}}

\setcopyright{acmcopyright}
\copyrightyear{2018}
\acmYear{2018}
\acmDOI{10.1145/1122445.1122456}

\begin{document}

\title{Measuring Equity: Funnel Representation Measurement}

\author{Wentao Su}
\affiliation{%
  \institution{LinkedIn Corporation}
  \city{Sunnyvale}
  \state{California}
  \country{USA}
}

\author{Guillaume Saint-Jacques}
\affiliation{%
  \city{Sunnyvale}
  \state{California}
  \country{USA}
}


\begin{abstract}
We present a methodology to measure the gender representation for online product funnels. It is a part of the overall equity framework to better understand our products through funnel analysis. By leveraging the coarsened exact matching method from causal inference literature, we show that the funnel survival ratio metric we design can detect the representation differences inherent in our products. Understanding how big the representation differences are between different member groups, as well as understanding what explains them, is critical for fostering more equitable outcomes. 
\end{abstract}




\maketitle
\pagestyle{empty}

\section{Introduction}
As web and social media companies are under intense scrutiny about the fairness issues related to their products, researchers and developers are focusing more on designing new methods that are more inclusive and transparent. This paper is one of such efforts to design a methodology to measure the representation in an equity framework through funnel analysis and have a data-driven impact on equity to help us track the equity progress.

When it comes to web and social media products, two main specificities need to be considered. First, they often follow a product funnel structure, starting from user registration, all the way to obtaining a certain goal or value, such as purchasing a merchant or getting a job. For example, a job seeker might join a career website, create a profile, search for a job, look through AI-recommended jobs, apply, hear back, and finally get hired. Second, they involve diverse members, with different careers and qualifications, and who come to the website for potentially different purposes at different times in their lives. It is important to analyze the fairness from the holistic view of the product funnels so that we are able to detect bias and differences for certain group of people down through the funnel.

We set out to build a tool that would allow assessment and detection of large representation differences, taking the above two specificities into account. There are already numerous product funnel analysis or metric designs in the market where people can calculate funnel conversion ratio for different business contexts \cite{chiong2020understanding}. However, there is not much discussion on funnel representation analysis, which is to design funnel metrics in an equity framework to measure overall representation for product funnels. Understanding how big the representation differences are between different member groups, as well as understanding what explains them, is critical for fostering more equitable outcomes for the digital products.

First, we provide an overview of the design of funnel representation metrics and discuss what the funnel representation measurement offers. Then we explains the methodology details including the coarsened exact matching (CEM) and p-value calculation in the context of A/B experiments. We conclude in the final section.

\section{What Does Funnel Representation Measurement Do? }
We first create a funnel with funnel layers defined by the product owners like the job seeker funnel mentioned above. Since we want to measure the representation of member groups through the funnels, the next variable of interest is the type of group representation we would like to gauge and it could be gender, ethnicity, age, or combinations of them. Our first use case involves grouping members by gender (male and female), but this approach can be extended to any number of groups in which cases we can compare the funnel survival ratio for every pair of groups or based on user’s interest. So far representation measurement based on binary gender data is conducted due to data availability and completeness.

First, we calculate the raw count of female and male for each layer of funnel and design four main demographic parity metrics that we use to measure representation and offer insights on what may explain unexplained differences for users.
\begin{table*}
  \caption{contributor funnel table with equity metrics }
  \label{tab:contributor funnel}
  \begin{tabular}{cp{2.5cm}cccp{2cm}p{2cm}p{3cm}}
    \toprule
    Funnel Step & Funnel Event& Female obs & Male obs & Raw Ratio & Normalized \newline Ratio &  Funnel Survival Ratio & Adjusted Funnel  \newline Survival Ratio \\
    \midrule
    1 & Active Users & 100M & 150M & 66.7\% & 1 & n.a. &n.a.  \\
    2 & Contributors & 15M & 25M & 60\% & 90\% & 90\% & 95\% (CI 94.2\%-94.8\%) \\
    3 & Contributors \newline with Response & 5M & 10M & 50\% & 75\%& 83.3\%& 90\% (CI 89\%-91\%)  \\

    \bottomrule
  \end{tabular}
  \footnotesize\emph{Note:} All the numbers in the table are hypothetical.
  \end{table*}

\begin{table*}
  \caption{contributor funnel table with equity metrics}
  \label{tab:metrics}
  \begin{tabular}{p{2.8cm}p{7.5cm}p{6cm}}
    \toprule
    Metric Name & Definition & Examples\\
    \midrule
    Raw Ratio &  \(\displaystyle \frac{Total\ outcomes\ for\ the\ lowest\ performing\ category}{Total\ outcomes\ for\ the\ highest\ performing\ category}  \)  & Number of sessions by women/number of sessions by men \\
    Normalized Ratio &  \(\displaystyle \frac{Mean\ outcomes\ for\ the\ lowest\ performing\ category}{Mean\ outcomes\ for\ the\ highest\ performing\ category}  \) &Mean sessions by non-deep dormant women / mean sessions by non-deep dormant men \\
    Funnel Survival Ratio &  \(\displaystyle \frac{Normalized\ ratio\ at\ the\ current\ funnel\ step}{Normalized\ ratio\ at\ the\ previous\ funnel\ step}   \)& Normalized ratio at job applications / Normalized ratio at job views \\
    Adjusted Funnel \newline Survival Ratio &  \(\displaystyle \frac{Controlled\ normalized\ ratio\ at\ the\ current\ funnel\ step}{Controlled\ normalized\ ratio\ at\ the\ previous\ funnel\ step} \)& Same as above, controlling for confounders like industry, country, etc. \\
    \bottomrule
  \end{tabular}
\end{table*}

Table~\ref{tab:contributor funnel} shows a use case of the funnel representation measurement for the member contributor funnel in a social media platform, which is defined as the active monthly users who both contribute to the network by writing a post and receive a response after the contribution. We will have a detailed description of these metrics in Technical Discussion section but the idea is that for every funnel composed of n layers, n - 1 survival ratios are derived so that users know which two layers down the funnel have representation issue.

\begin{figure}[h]
  \centering
  \includegraphics[width=\linewidth]{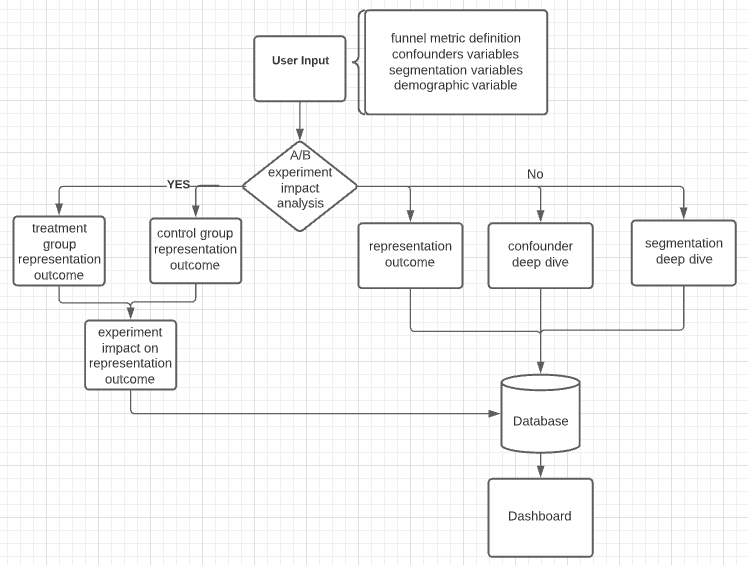}
  \caption{flow chart for funnel representation measurement}
  \Description{this is the flow chart}  \label{fig:flow}
\end{figure}

The tool requires only a few inputs from the users: the funnel metric definition, the population base and external variables to be controlled, as well as the type of groups we want to measure. Based on the input, the tool computes the metrics and generates the representation outcomes as well as by each segment if selected by users. If users provide additional information on an A/B experiment, the product can measure the impact of that experiment on representation metrics. Figure \ref{fig:flow} shows the measurement process.
Currently this representation measure tool has been actively used in several pillar products and is 100\% self-service for the extension to different use cases.

\section{Technical Discussion}
\subsection{Four Representation Metrics}
We define four metrics to measure representation from different perspectives.

There are 4 metrics that we design to measure the status of representation for product funnels: raw ratio, normalized ratio, funnel survival ratio, and adjusted funnel survival ratio shown in Table~\ref{tab:metrics}.  In gender male-female context, the raw ratio is the female over male ratio of raw metric count in each funnel layer; the normalized ratio is the normalized raw ratio based on each funnel base population; the funnel survival ratio is the ratio of two normalized ratios from downstream layer to upstream layer, which can be understood as raw gender survival ratio. 

\begin{figure}[h]
  \centering
  \includegraphics[width=\linewidth]{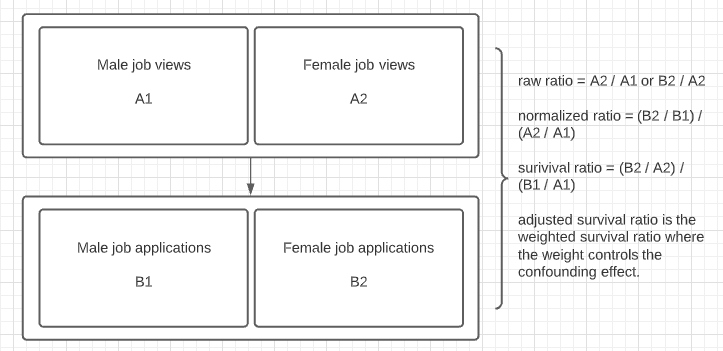}
  \caption{an example to illustrate the metric definition}
  \Description{this is the metric definition chart}  
  \label{fig:metric definition}
\end{figure}

The last metric, adjusted funnel survival ratio, is the funnel survival ratio adjusted for the confounding effect and our preferred metric for representation measurement. For many digital product funnels, the representation outcomes are influenced by confounders such as job functions, industries, countries - member attributes external to web and social media companies. We want to get the effect of group categories on representation outcome after taking out those external factors so that we can show the product representation status attributed to the internal platform.  As an example, we have seen more gender skewness in terms of member engagement in some industries than others and we want to control the effect of this external factor on the funnel survival ratio.  We leverage a popular causal inference method - coarsened exact matching (CEM) to adjust the confounding effect, discussed in section \ref{cem}. Essentially, confounders are external factors that would contribute to explain the differences between groups. Figure \ref{fig:metric definition} uses a job view two-layer funnel to further illustrate the metric definition.
\subsection{Color-coding Rule}

The thresholds are product-specific and users should determine their own threshold based on each product. As a starting point, we recommend three-level status for monitoring representation of each product funnel based on principles of Equal Employment Opportunity Commission (EEOC) rules. \footnote{See official EEOC document \url{https://www.eeoc.gov/laws/guidance/questions-and-answers-clarify-and-provide-common-interpretation-uniform-guidelines} in detail.}

\begin{itemize}
  \item Green -  funnel survival ratio change is under the 1-5\% range - indicates healthy representation
  \item Yellow - funnel survival ratio change is in 5-25\% range - issues a warning 
  \item Red - funnel survival ratio change over 10-25\% change - requires deep-dive analysis 
\end{itemize}

\begin{table}
  \caption{color-coding rule table with three scenarios}
  \label{tab:color-coding}
  \begin{tabular}{cccc}
    \toprule
    Status&Strict&Middle & Loose\\
    \midrule
    Green & <1\% &  <3\% & <5\%\\
    Yellow & <1-10\% &  <3-15\% & 5-20\%\\
    Red & >10\% &  <>15\% & >20\%\\
  \bottomrule
\end{tabular}
\end{table}

The exact value can be fixed empirically in terms of the percentage deviation of the adjusted survival ratio from the parity 1, as shown in Table~\ref{tab:color-coding}. The more confident we are that we capture all confounders, the less the impact these confounders have on the adjusted survival ratio, in which case the lower thresholds are recommended, as the result from the analysis is robust. On the other hand, if the list of confounders is not complete or with poor quality, a larger buffer is needed to take that into consideration.

\subsection{CEM Methodology For Confounder Adjustment}
\subsubsection{Why Do We Need To Adjust Confounders?} 

Web and social media companies can have a very diverse member of population from different background. Many of these factors are external to the product funnels and may be unevenly distributed among demographic categories, and thus these factors or confounders can heavily influence the representation outcomes for online products. In order to address this issue, we need to control for the representation distribution of factors outside the scope of the companies and get a true understanding of representation status intrinsic to these products. It should be noted again that the accuracy of the confounder adjustment heavily depends on the exhaustive list of the confounders that users can find for each product and therefore the thresholds of representation metrics are product-specific as mentioned above.

\subsubsection{CEM Methodology} 
\label{cem}
The basic idea of CEM is to coarsen each variable by recoding so that substantively indistinguishable values are grouped and assigned the same numerical value \cite{iacus2012causal}. Then, the ‘‘exact matching’’ algorithm is applied to the coarsened data to determine the matches and to prune unmatched units. Finally, the coarsened data are discarded and the original uncoarsened values of the matched data are retained.

After the coarsening, the CEM algorithm creates a set of strata, say $s \in S$, each with the same coarsened values of $X$. Strata that contain at least one treated and one control unit or data observation - male and female based on this discussion - are retained; units in the remaining strata are removed from this sample. We denote by $ \tau ^s $ the treated units in stratum $s$ and by $m_T^s  = count\ of\ \tau ^s$ the number of treated units in the stratum, similarly for the control units, that is $C^s$ and $m_C^s = count\ of\ C^s$ The number of matched units in which case there are both treated and control units in every strata are respectively for treated and controls, $m_T =  \cup_{ s \in S} m_T^s $ and  $m_C =  \cup_{ s \in S} m_C^s $.  To each matched unit $i$ in stratum $s$, CEM assigns the following weights:
 \begin{equation}
    w_{i} =
    \begin{cases}
      \frac{m_{T}}{m_{T}^s} *  \frac{m^s}{m}, & i \in T^s \\
      \frac{m_{C}}{m_{C}^s} *  \frac{m^s}{m}, & i \in C^s
    \end{cases}
  \end{equation}

Unmatched units receive weight $w_i = 0$.

CEM therefore assigns to matching the task of eliminating all imbalances (i.e., differences between the treated and control groups) beyond some chosen level defined by the coarsening. 
From the matched data, we can calculate the funnel survival ratio of each strata $Y_T$ and incorporate the weights to adjust for the confounding effect. The weighted sum of each strata's funnel survival ratio is the final adjusted survival ratio of the whole funnel.
 \begin{equation}
Adjusted\ Funnel\ survival\ ratio =  \frac{\frac{\sum(Y_T * W_T)}{\sum W_T}}{\frac{\sum(Y_C * W_C)}{\sum W_C}}
  \end{equation}

\begin{table*}
  \caption{the output example of representation measurement of an A/B experiment}
  \label{tab:feed funnel}
  \begin{tabular}{ccccccc}
    \toprule
         && Baseline Model &  &   &  Treatment Model &  \\
        Funnel Event & Female obs & Male obs & Raw Ratio & Female obs & Male obs & Raw Ratio \\
    \midrule
    Feed Viewers & 22M & 34M & 64.2\% & 22M & 35M & 64.1\%  \\
    Daily Unique Contributors(DUC) & 10M & 16M & 61.4\% & 10M & 16M & 61.3\%  \\
    DUC percentage & 45\%  & 47\% & 95.6\% & 45\%(0.15\%, 0.23) & 47\%(0.15\%, 0.09)& 95.6\%  \\
    Adjusted Funnel Survival Ratio & & & 97.6\% & & & \textbf{97.5\%(-0.1\%, 0.55)} \\
    \bottomrule
  \end{tabular}
  \footnotesize\emph{Note:} All the numbers in the table are hypothetical. Two numbers within hypotheses are lift of the metric and its p-value.
\end{table*}

\subsubsection{P-value For Metric Difference in A/B Experiments}
The funnel representation metrics can also be applied to A/B experiments context. When a new feature or a new model is launched to the public, we not only want to know how the true north business metrics respond and whether the responses are significant, increasingly we are also interested in how the feature or model impacts the representation metrics in certain categories like whether a new machine learning model changes the equity representation in one product funnel.
To answer this question, we can leverage the property of the survival ratio formula and derive the variance of the different ratios between the treatment and control groups. Below is the calculation process:

\begin{enumerate}
  \item Calculate the funnel survival ratio for the treatment group and control group separately.
  \item Obtain confidence interval for the ratio of two survival ratios \cite{katz1978obtaining}. The asymptotic variance of the logarithm of two ratios are $\frac{1-p}{pm} + \frac{1-q}{qm}$ where $p$ and $q$ are survival rates of two groups (in this case females and males) and $n$ and $m$ are sample sizes of two groups so that 
  
   \begin{equation}
   variance(log(SR)) = variance(log(\frac{SR_f}{SR_m}) = \frac{1-p}{pm} + \frac{1-q}{qm}
  \end{equation}
  Suppose $SR_t$ and $SR_c$ represent the survival ratios for treatment and control group respectively, the variance of their ratio is the sum of variance of the logarithm of the ratio from which we can get above.   
     \begin{align}
   variance(log(\frac{SR_t}{SR_c})) & = variance(log(\frac{SR_t}{SR_c}) \\
                                                      & = variance(log(SR_t)) + variance(log(SR_c))
  \end{align}
  \item Calculate the p-value for the significance of survival ratio difference in treatment and control group

     \begin{equation}
  z\ score = \frac{log(\frac{SR_t}{SR_c}) - 0}{variance(log(\frac{SR_t}{SR_c}))}
  \end{equation}
\end{enumerate}

The output from this tool looks like Table~\ref{tab:feed funnel} where the experiment tests a new version of the feed model versus the old version. The key metric of interest is the adjusted funnel survival ratios at the bottom right with the lift of two ratios and the lift’s p-value within the parentheses. In this example, although we see a 0.1\% drop in adjusted survival ratio meaning that the female feed viewers are put in a slightly disadvantageous position for new model compared to the old one, this result is not statistically significant.

\section{Conclusion}

This paper discusses the representation tool we built to measure the overall representation for product funnels. The unique feature compared to traditional funnel analysis is that funnel metrics are designed in an equity framework and causal inference methodology is leveraged into the design of representation metrics in order to control the external factors.

With the help of this analytical tool, we now have a better understanding of the representation gap for web and social media products. Future research includes how to create an equity measurement platform where multiple fairness detection and mitigation tools are utilized together to empower us to foster more equitable outcomes for the product users.

\begin{acks}
We would like to express our sincere thanks to Yinyin Yu, Rina Friedberg, and Karthik Rajkumar for their helpful comments and feedback, 
and Parvez Ahammad for his support throughout this project.
\end{acks}

\newpage 

\thispagestyle{empty}
\bibliographystyle{ACM-Reference-Format}
\bibliography{sample-base}

%
%
%
%
%
%
%
%

\end{document}